\documentclass[12pt]{article}
\usepackage{graphicx}
\begin{document}

\title{{\Large NP PROBLEM\ IN\ QUANTUM\ ALGORITHM}}
\author{Masanori Ohya and Natsuki Masuda \\
Science University of Tokyo\\
Noda City, Chiba 278-8510, JAPAN}
\date{}
\maketitle

\begin{abstract}
In complexity theory, there exists a famous unsolved problem whether NP can
be P or not. In this paper, we discuss this aspect in SAT (satisfiability)
problem, and it is shown that the SAT can be solved in plynomial time by
means of quantum algorithm.
\end{abstract}

\section{Introduction}

Although the ability of computer is highly progressed, there are several
problems which may not be solved effectively, namely, in polynomial time.
Among such problems, NP problem and NP complete problem are fundamental. It
is known\cite{GJ} that all NP complete problems are equivalent and an
essential question is whether there exists an algorithm to solve a NP
complete problem in polynomial time.

After pioneering works of Feymann\cite{Fey} and Deutsch\cite{Deu}, several
important works have been done on quantum algorithms by Deutsch and Josa\cite
{DJ}, Shor\cite{Sho}, Ekert and Josa\cite{EJ} and many others\cite{O1}. The
computation in quantum computer is performed on a tensor product Hilbert
space, and its fundamental point is to use quantum coherence of states. All
mathematical features of quantum computer and computation are summarized in
\cite{O1}.

In this paper, we discuss the quantum algorithm of the SAT problem and point
out that this problem, hence all other NP problems, can be solved in
polynomial time by quantum computer if the superposition of two orthogonal
vectors $\left| 0\right\rangle $ and $\left| 1\right\rangle $ is physically
detected.

\section{SAT Problem}

Let $X\equiv \left\{ x_{1},\cdots ,x_{n}\right\} $ be a set. Then $x_{k}$
and its negation $\overline{x}_{k}$ $(k=1,2,\cdots ,n)$ are called literals
and the set of all such literals is denoted by \\
$X^{^{\prime }}=\left\{ x_{1},%
\overline{x}_{1},\cdots ,x_{n},\overline{x}_{n}\right\} $. The set of all
subsets of $X^{^{\prime }}$ is denoted by $\mathcal{F}(X^{^{\prime }})$ and
an element $C\in \mathcal{F}(X^{^{\prime }})$ is called a clause. We take a
truth assignment to all variables $x_{k}.$ If we can assign the truth value
to at least one element of $C,$ then $C$ is called satisfiable. When $C$ is
satisfiable, the truth value $t(C)$ of $C$ is regarded as true, otherwise,
that of $C$ is false. Take the truth values as ''true$\leftrightarrow 1,$
false$\leftrightarrow 0".$ Then

\[
C\mbox{ is satisfiable iff }t(C)=1.
\]

Let $L=\left\{ 0,1\right\} $ be a Boolean lattice with usual join $\vee $
and meet $\wedge ,$ and $t(x)$ be the truth value of a literal $x$ in $X.$
Then the truth value of a clause $C$ is written as $t(C)\equiv \vee _{x\in
C}t(x).$ Moreover the set $\mathcal{C}$ of all clauses $C_{j}$ $%
(j=1,2,\cdots m)$ is called satisfiable iff the meet of all truth values of $
C_{j}$ is 1; $t(\mathcal{C})$ $\equiv \wedge _{j=1}^{m}t(C_{j})=1.$ Thus the
SAT problem is written as follows:

\bigskip

\noindent
{\bf Definition 1.}
{\it
SAT Problem: Given a set $X\equiv \left\{ x_{1},\cdots ,x_{n}\right\} $ and
a set $\mathcal{C=}\left\{ C_{1},C_{2},\cdots ,C_{m}\right\} $ of clauses,
determine whether $\mathcal{C}$ is satisfiable or not.
}

\bigskip

That is, this problem is to ask whether there exsits a truth assignment to
make $\mathcal{C}$ satisfiable.

It is known\cite{GJ} in usual algorithm that it is polynomial time to check
the satisfiability only when a specific truth assignment is given, but we
can not determine the satisfiability in polynomial time when an assignment
is not specified.

\section{Quantum Algorithm of SAT}

Let 0 and 1 of the Boolean lattice $L$ be denoted by the vectors $\left|
0\right\rangle \equiv \left(
\begin{array}{l}
1 \\
0
\end{array}
\right) $ and $\left| 1\right\rangle \equiv \left(
\begin{array}{l}
0 \\
1
\end{array}
\right) $ in the Hilbert space ${\bf C}^{2}$, respectively. That is, the
vector $\left| 0\right\rangle $ corresponds to falseness and $\left|
1\right\rangle $ does to truth.

As we explained in the previous section, an element $x\in X$ can be denoted
by 0 or 1, so by $\left| 0\right\rangle $ or $\left| 1\right\rangle .$ In
order to describe a clause $C$ with at most n length by a quantum state, we
need the n-tuple tensor product Hilbert space $\mathcal{H\equiv }$ $\otimes
_{1}^{n}$${\bf C}^{2}$. For instance, in the case of $n=2$, given $
C=\left\{ x_{1},x_{2}\right\} $ with an assignment $x_{1}=0$ and $x_{2}=1,$
then the corresponding quantum state vector is $\left| 0\right\rangle
\otimes \left| 1\right\rangle ,$ so that the quantum state vector describing
$C$ is generally written by $\left| C\right\rangle =\left|
x_{1}\right\rangle \otimes \left| x_{2}\right\rangle \in $ $\mathcal{H}$
with $x_{k}=0$ or $1$ (k=1,2).

The quantum computation is performed by a unitary gate constructed from
several fundamental gates such as Not gate, Controlled-Not gate, \\
Controlled-Controlled Not gate \cite{EJ,O1}. Once $X\equiv \left\{
x_{1},\cdots ,x_{n}\right\} $ and \\
$\mathcal{C=}\left\{ C_{1},C_{2},\cdots
,C_{m}\right\} $ are given, the SAT is to find the vector $\left| f\left(
\mathcal{C}\right) \right\rangle \equiv $ $\wedge _{j=1}^{m}\vee _{x\in
C_{j}}t(x),$ where $t(x)$ is $\left| 0\right\rangle \ $or $\left|
1\right\rangle $ when $x=0$ or 1, respectively, and $t(x)\wedge t(y)\equiv
t(x\wedge y)$, $t(x)\vee t(y)\equiv t(x\vee y).$

We consider the quantum algorithm for the SAT problem. Since we have $n$
variables $x_{k}\left( k=1,\cdots n\right) $ and a quantum computation
produces some dust bits, the assignments of the $n$ variables and the dusts
are represented by $n$ qubits and $l$ qubits in the Hilbert space $\otimes
_{1}^{n}$${\bf C}^{2}\otimes _{1}^{l}{\bf C}^{2}$. Moreover the
resulting state vector $\left| f\left( \mathcal{C}\right) \right\rangle $
should be added, so that the total Hilbert space is

\[
\mathcal{H\equiv }\otimes _{1}^{n}\mathbf{C}^{2}\otimes _{1}^{l}\mathbf{C}%
^{2}\otimes \mathbf{C}^{2}.
\]

Let us start the quantum computation of SAT\ problem from an initial vector $%
\left| v_{0}\right\rangle \equiv \otimes _{1}^{n}\left| 0\right\rangle
\otimes _{1}^{l}\left| 0\right\rangle \otimes \left| 0\right\rangle $ when $%
\mathcal{C}$ contains n Boolean variables $x_{1},\cdots x_{n}.$ We apply the
discrete Fourier transformation denoted by $U_{F}\equiv \otimes _{1}^{n}%
\frac{1}{\sqrt{2}}\left(
\begin{array}{ll}
1 & 1 \\
1 & -1
\end{array}
\right) $ to the part of the Boolean variables of the vector $\left|
v_{0}\right\rangle ,\ $then the resulting state vector becomes

\[
\left| v\right\rangle \equiv U_{F}\otimes _{1}^{l}I\left| v_{0}\right\rangle
=\frac{1}{\sqrt{2^{n}}}\otimes _{1}^{n}\left( \left| 0\right\rangle +\left|
1\right\rangle \right) \otimes _{1}^{l}\left| 0\right\rangle \otimes \left|
0\right\rangle ,
\]
where $I$ is the identity matrix in $\mathbf{C}^{2}.$ This vector can be
written as

\[
\left| v\right\rangle =\frac{1}{\sqrt{2^{n}}}\sum_{x_{1},\cdots
,x_{n}=0}^{1}\otimes _{j=1}^{n}\left| x_{j}\right\rangle \otimes
_{1}^{l}\left| 0\right\rangle \otimes \left| 0\right\rangle .
\]

Now, we perform the quantum computer to check the satisfiability, which will
be done by a unitary operator $U_{f}$ properly constructed by unitary gates.
Then after the computation by $U_{f}$, the vector $\left| v\right\rangle $
goes to

\begin{eqnarray*}
\left| v_{f}\right\rangle &\equiv &U_{f}\left| v\right\rangle =\frac{1}{%
\sqrt{2^{n}}}\sum_{x_{1},\cdots ,x_{n}=0}^{1}U_{f}\otimes _{j=1}^{n}\left|
x_{j}\right\rangle \otimes _{1}^{l}\left| 0\right\rangle \otimes \left|
0\right\rangle \\
&=&\frac{1}{\sqrt{2^{n}}}\sum_{x_{1},\cdots ,x_{n}=0}^{1}\otimes
_{j=1}^{n}\left| x_{j}\right\rangle \otimes _{i=1}^{l}\left|
y_{i}\right\rangle \otimes \left| f\left( x_{1},\cdots ,x_{n}\right)
\right\rangle ,
\end{eqnarray*}
where $f\left( x_{1},\cdots ,x_{n}\right) \equiv f\left( \mathcal{C}\right) $
because $\mathcal{C}$ contains $x_{1},\cdots x_{n},$ and $\left|
y_{i}\right\rangle $ are the dust bits produced by the computation. As we
will explain in an example below, the unitary operator $U_{f}$ is concretely
constructed.

The final step to check the satisfiability of $\mathcal{C}$ is to apply the
projection $E\equiv \otimes _{1}^{n+l}I\otimes \left| 1\right\rangle
\left\langle 1\right| $ to the state $\left| v_{f}\right\rangle ,$
mathematically equivalent , to compute the value $\left\langle v_{f}\right|
E\left| v_{f}\right\rangle .$ If the vector $E\left| v_{f}\right\rangle $
exists or the value $\left\langle v_{f}\right| E\left| v_{f}\right\rangle $
is not 0, then we conclude that $\mathcal{C}$ is satisfiable.The value of $
\left\langle v_{f}\right| E\left| v_{f}\right\rangle $ corresponds to that
of the ramdom algorithm as we will see in an example of the next section and
it may or may not be obtained in polynominal time. Let us consider an
operator $V_{\theta },$ given by
\[
V_{\theta }\equiv \otimes _{1}^{n+l}(A\left| 0\right\rangle \left\langle
0\right| +B\left| 1\right\rangle \left\langle 1\right| )\otimes e^{i\theta
f\left( \mathcal{C}\right) }I,
\]
and apply it to the vector $\left| v_{f}\right\rangle ,$ where $A\equiv
\frac{1}{\sqrt{2}}\left(
\begin{array}{ll}
1 & 1 \\
1 & -1
\end{array}
\right) $ and $B\equiv \frac{1}{\sqrt{2}}\left(
\begin{array}{ll}
1 & 1 \\
-1 & 1
\end{array}
\right) $\ \ and $\theta $\ is a certain constant describing the phase of
the vector $\left| f\left( \mathcal{C}\right) \right\rangle .$ The resulting
vector is the superposition of two vectors with some constants $\alpha ,$ $
\beta $ such as

\[
V_{\theta }\left| v_{f}\right\rangle =\frac{1}{\sqrt{2^{n+l}}}\otimes
_{1}^{n+l}\left( \left| 0\right\rangle +\left| 1\right\rangle \right)
\otimes \left( \alpha \left| 0\right\rangle +\beta e^{i\theta }\left|
1\right\rangle \right) ,
\]
one of which is polarized with $\theta $ and another is non-polarized. The
existence of the mixture of two vectors $\left| 0\right\rangle $ and $
e^{i\theta }\left| 1\right\rangle $ is a starting point of quantum
computation, which implies the satisfiability.

\section{Example of Computation}

Let us explain the quantum computation for SAT in tha case $%
X=\{x_{1},x_{2},x_{3}\}$ and $\mathcal{C}=\left\{ \left\{ x_{1}\right\}
,\left\{ x_{2},x_{3}\right\} ,\left\{ x_{1},\overline{x}_{3}\right\}
,\left\{ \overline{x}_{1},\overline{x}_{2},x_{3}\right\} \right\} .$ The
resulting state \\
$\left| f\left( x_{1},x_{2},x_{3}\right) \right\rangle $ is
written as

\[
\left| f\left( x_{1},x_{2},x_{3}\right) \right\rangle =\left|
x_{1}\right\rangle \wedge \left( \left| x_{2}\right\rangle \vee \left|
x_{3}\right\rangle \right) \wedge \left( \left| x_{1}\right\rangle \vee
\left| \overline{x}_{3}\right\rangle \right) \wedge \left( \left| \overline{x%
}_{1}\right\rangle \vee \left| \overline{x}_{2}\right\rangle \vee \left|
x_{3}\right\rangle \right)
\]
In the quantum computation, it is not necessary to substitute all values of $
x_{j}$ $\left( j=1,2,3\right) $ as the classical computation, we have only
to use a unitary operator $U_{f}$ for the computation of $\left| f\left(
x_{1},x_{2},x_{3}\right) \right\rangle .$ This unitary operator $U_{f}$ is
constructed as follows: Let $U_{NOT},$ $U_{CN}\ $and $U_{CCN}$ be Not gate
on ${\bf C}^{2}$, Controlled-Not gate on ${\bf C}^{2}\otimes {\bf C}^{2}$ and
Controlled-Controlled-Not gate on ${\bf C}^{2} \otimes {\bf C}^{2} \otimes
 {\bf C}^{2}$, respectively, which are given by

\[
\begin{array}{l}
U_{NOT}=\left| 0\right\rangle \left\langle 1\right| +\left| 1\right\rangle
\left\langle 0\right| \\
U_{CN}=\left| 0\right\rangle \left\langle 0\right| \otimes I+\left|
1\right\rangle \left\langle 1\right| \otimes \left( \left| 0\right\rangle
\left\langle 1\right| +\left| 1\right\rangle \left\langle 0\right| \right)
\\
U_{CCN}=\left| 0\right\rangle \left\langle 0\right| \otimes \left|
0\right\rangle \left\langle 0\right| \otimes I+\left| 0\right\rangle
\left\langle 0\right| \otimes \left| 1\right\rangle \left\langle 1\right|
\otimes I+\left| 1\right\rangle \left\langle 1\right| \otimes (\left|
0\right\rangle \left\langle 0\right| \otimes \left| 1\right\rangle
\left\langle 0\right| )
\end{array}
\]
Then the unitary operator $U_{f}$ is determined by the combination of the
above three unitaries as

\[
U_{f}\equiv U_{36}U_{35}\cdots U_{2}U_{1},
\]
where, for instance,

\begin{eqnarray*}
U_{1} &\equiv &\left| 0\right\rangle \left\langle 0\right| \otimes
_{1}^{23}I+\left| 1\right\rangle \left\langle 1\right| \otimes
_{1}^{2}I\otimes \left( \left| 0\right\rangle \left\langle 1\right| +\left|
1\right\rangle \left\langle 0\right| \right) \otimes _{1}^{20}I \\
U_{2} &\equiv &I\otimes \left| 0\right\rangle \left\langle 0\right| \otimes
_{1}^{22}I+I\otimes \left| 1\right\rangle \left\langle 1\right| \otimes
_{1}^{2}I\otimes \left( \left| 0\right\rangle \left\langle 1\right| +\left|
1\right\rangle \left\langle 0\right| \right) \otimes _{1}^{19}I \\
U_{3} &\equiv &\otimes _{1}^{2}I\otimes \left| 0\right\rangle \left\langle
0\right| \otimes _{1}^{21}I+\otimes _{1}^{2}I\otimes \left| 1\right\rangle
\left\langle 1\right| \otimes _{1}^{2}I\otimes \left( \left| 0\right\rangle
\left\langle 1\right| +\left| 1\right\rangle \left\langle 0\right| \right)
\otimes _{1}^{18}I
\end{eqnarray*}
and other $U_{4},\cdots ,U_{36}$ are similarly constructed (see the
computation diagram 1). In this case, we need $20$ dust bits (the number of
the dust bits needed in a general case is counted in the next section), so
that $U_{f}$ is operated on the Hilbert space $\otimes _{1}^{24} {\bf C}^{2}$.

Starting from the initial vector $\left| v_{0}\right\rangle \equiv \otimes
_{1}^{3}\left| 0\right\rangle \otimes _{1}^{20}\left| 0\right\rangle \otimes
\left| 0\right\rangle ,$ the final vector is

\begin{eqnarray*}
\left| v_{f}\right\rangle &=&U_{36}\cdots U_{1}U_{F}\left|
v_{0}\right\rangle =U_{f}U_{F}\left| v_{0}\right\rangle \\
&=&\frac{1}{\sqrt{2^{3}}}\left( \left|
0,0,0,0,1,1,0,1,0,1,0,0,0,1,1,0,0,0,1,1,1,0,1;0\right\rangle \right. \\
&&+\left| 1,0,0,1,1,1,0,0,0,1,1,0,0,1,1,1,0,0,1,1,1,0,1;0\right\rangle \\
&&+\left| 0,1,0,0,0,1,1,1,0,1,0,1,0,1,1,0,1,0,1,1,1,0,1;0\right\rangle \\
&&+\left| 0,0,1,0,1,0,1,1,1,0,0,0,0,0,1,0,1,0,0,1,0,0,0;0\right\rangle \\
&&+\left| 1,1,0,1,0,1,1,0,0,1,1,1,1,1,0,1,1,1,1,1,0,1,0;0\right\rangle \\
&&+\left| 1,0,1,1,1,0,1,0,1,1,1,0,0,0,1,1,1,1,1,1,1,1,1;1\right\rangle \\
&&+\left| 0,1,1,0,0,0,1,1,1,0,0,1,0,0,1,0,1,0,0,1,0,0,0;0\right\rangle \\
&&\left. +\left|
1,1,1,1,0,0,1,0,1,1,1,1,1,0,1,1,1,1,1,1,1,1,1;1\right\rangle \right) ,
\end{eqnarray*}
where we used the notation

\[
\left| x_{1},x_{2},x_{3},y_{1},\cdots ,y_{20};f\left(
x_{1},x_{2},x_{3}\right) \right\rangle \equiv \otimes _{j=1}^{3}\left|
x_{j}\right\rangle \otimes _{i=1}^{20}\left| y_{i}\right\rangle \otimes
\left| f\left( x_{1},x_{2},x_{3}\right) \right\rangle
\]
Applying the operator $V_{\theta }$ of the previous section to the vector $%
\left| v_{f}\right\rangle ,$ we obtain

\[
\frac{1}{\sqrt{2^{23}}}\otimes _{1}^{23}\left( \left| 0\right\rangle +\left|
1\right\rangle \right) \otimes (\frac{\sqrt{3}}{2}\left| 0\right\rangle +%
\frac{1}{2}e^{i\theta }\left| 1\right\rangle ),
\]
which is the superposition of two vectors $\left| 0\right\rangle $ and the
polarized vector $\left| f\left( \mathcal{C}\right) \right\rangle =\left|
1\right\rangle .$ Moreover when we measure the operator $E\equiv \otimes
_{1}^{23}I\otimes \left| 1\right\rangle \left\langle 1\right| $ in the state
$V_{\theta }\left| v_{f}\right\rangle ,$ we obtain

\[
\left\langle v_{f}V_{\theta }\right| E\left| V_{\theta }v_{f}\right\rangle =%
\frac{1}{4}.
\]
These conclude the satisfiability of $\mathcal{C}$.

\begin{center}
\includegraphics[width=9cm,height=12cm]
{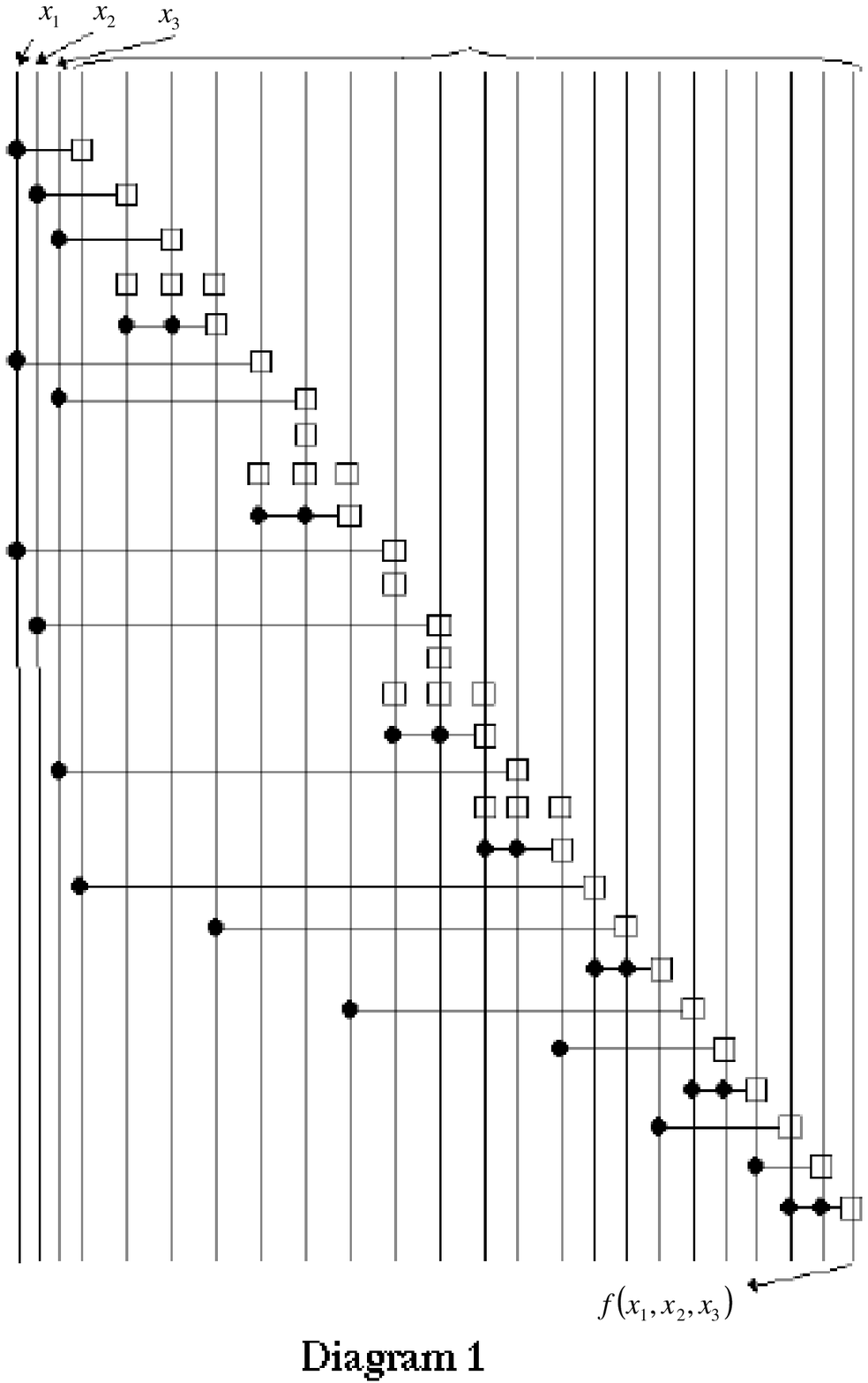}
\end{center}

\section{Complexity of Quantum Algorithm for SAT}

Here we discuss the number of steps for the quantum algorithm of the SAT
problem. The size of input with n variables $x_{k}$ and their negations $%
\overline{x}_{k}$ $\left( k=1,\cdots ,n\right) $and also $\mathcal{C}%
=\{C_{1},\cdots ,C_{m}\}$ is $N_{1}=\log n+2mn$ like the classical algorithm
because $\left| C_{k}\right| ,$ the number of the elements in $C_{k},$ is at
most $n$, so that it is the polynominal order $(O(mn)).$ The number of the
dust bits to compute $f\left( \mathcal{C}\right) $ is caused from that of
the operations and substitutions of AND and OR, so that the maximum number
(complexity) $N_{2}$ of the dust bits needed is the same as that of the
classical case, namely, $N_{2}=$ (the numbers of AND and OR operations )$-$
(the steps to take the negation) = ($5mn-1)-mn=4mn-1.$

The steps $N_{3}$ needed to obtain the vector $\left| f\left( \mathcal{C}
\right) \right\rangle $ is counted as follows: Fisrt take $1$ step for the
discrete Fourier transform to get the entangled vector, and we need $3mn$
steps for truth assignment and substitution to $n$ variables. Secondly to
compute the logical sum in each clause and take their logical product, the
complexity corresponding to the logical sum, whose gate is made from four
unitaries, is $4m(2n-1)$ and that corresponding to the logical product is $
m-1.$ Thus $N_{3}=1+3mn+4m(2n-1)+m-1$ $=11mn-3m.$

Finally to check the satisfiability of $\mathcal{C}$, we have to look at the
situation of $\left| f(\mathcal{C)}\right\rangle $ registered at the last
position of the tensor product state such as the last position of physical
registers (e.g., spins) lined up.\ This can be easily done by applying the
operator $V_{\theta }$ to $\left| v_{f}\right\rangle ,$ and the resulting
vector is a superposition of two vectors $\left| 0\right\rangle $ and $
e^{i\theta }\left| 1\right\rangle .$ We can obtain in polynominal time (at
most $n+l+1)$ the vector $\left| V_{\theta }v_{f}\right\rangle $ and the
value of $<V_{\theta }v_{f},EV_{\theta }v_{f}\mathcal{>=}\left| \beta
\right| ^{2}$ for the projection $E$ if needed. The existence of the above
superposition is a starting point of quantum computation, so that it should
be physically detected being different from both $\left| 0\right\rangle $
and $\left| 1\right\rangle $ , which implies the satisfiability. Thus the
quantum algorithm of the SAT problem is in polynominal order.

\bigskip

\noindent
{\large\bf Acknowledgment:} One of the authors (MO) appreciate Professor A.
Ekert for his critical reading and valuable comments.

\end{document}